\documentclass[twoside,12pt]{article}
\usepackage{epsfig}

\newcommand{\be}{\begin{equation}}
\newcommand{\ee}{\end{equation}}
\newcommand{\bea}{\begin{eqnarray}}
\newcommand{\eea}{\end{eqnarray}}

\begin{document}


\title{GAMOW-TELLER STRENGTH DISTRIBUTIONS for $\beta\beta$-DECAYING NUCLEI WITHIN CONTINUUM-QRPA}
\author{S.Yu. Igashov$^{\,1}$, V.A. Rodin$^{\,2}$, M.H. Urin$^{\,1,3}$, A. Faessler$^{\,2}$
\\
$^1$ Moscow Engineering Physics Institute (State University), Russia\\
$^2$Institute for Theoretical Physics, University of Tuebingen, Germany\\
$^3$Kernfysisch Versneller Instituut, University of Groningen, The Netherlands}

\maketitle

\begin{abstract} 
A version of the pn-continuum-QRPA is outlined and applied to describe the 
Gamow-Teller strength distributions for $\beta\beta$-decaying open-shell nuclei. 
The calculation results obtained for the pairs of nuclei $^{116}$Cd-Sn and 
$^{130}$Te-Xe are compared with available experimental data.

\end{abstract}

\section{Introduction}

Description of weak interaction in nuclei 
is often a challenge for models of nuclear structure. 
Numerous calculations of the nuclear $\beta\beta$-decay amplitudes 
well illustrate this statement (see, e.g. Refs. \cite{rev} and references therein). 
Uncertainties in theoretical calculations of the Gamow-Teller (GT) 
$2\nu\beta\beta$-decay amplitude $M^{2\nu}_{GT}$ have stimulated experimental 
studies of the GT$^{(\mp)}$-strengths of the $1^{+}$ states virtually excited 
in the decay process (see, e.g. Refs.~\cite{sas07,rak05}). 

As a double charge-exchange process, $2\nu(0\nu)\beta\beta$-decay is enhanced 
by nucleon pairing which is due to the singlet part of the particle-particle (p-p) 
interaction. 
The discrete quasiboson version of the quasiparticle RPA (pn-dQRPA) which accounts for the nucleon pairing 
is usually applied to calculate the $\beta\beta$-decay amplitudes in open-shell nuclei \cite{rev}. 
In spite of differences in model parameterizations of the nuclear mean field and residual interaction in 
the particle-hole (p-h) and p-p channels, all pn-dQRPA calculations reveal
marked sensitivity of 
the amplitude $M^{2\nu}_{GT}$ 
to the ratio $g_{pp}$ of the triplet to singlet strength of the p-p interaction.  
Physical reasons for such a general feature of all calculations were analyzed in Ref.~\cite{rod05}, where 
they were attributed to violation of the spin-isospin SU(4) symmetry  in nuclei. 
An identity transformation of the amplitude into sum of two 
terms was used in Ref.~\cite{rod05}. One term, which is due to the
p-p interaction only, depends linearly on $g_{pp}$ and vanishes at $g_{pp}=1$ when
the SU(4)-symmetry is restored in the p-p sector of a model Hamiltonian.
The second term is a smoother function of $g_{pp}$ at $g_{pp}\sim 1$, 
but exhibits a quadratic dependence on the strength of the mean-field spin-orbit term, 
which is the main source of violation of the spin-isospin SU(4)-symmetry in nuclei.

Understanding of general properties of the amplitude $M^{2\nu}_{GT}$ helps
to improve reliability of evaluation of $\beta\beta$-decay amplitudes.
For a quantitative analysis, we use here an isospin-selfconsistent pn-continuum-QRPA (pn-cQRPA) approach 
of Ref. \cite{rod03}, where this approach was applied to evaluate the GT$^{(\mp)}$ 
strength distributions in single-open-shell nuclei. 
In the reference the full basis of the single-particle (s-p) 
states was used in the p-h channel along with the Landau-Migdal forces, while the nucleon pairing was described 
within the simplest version of the BCS-model based on discrete basis of s-p states. 
A rather old version of the phenomenological isoscalar nuclear mean field 
(including the spin-orbit term) was used in Refs.~\cite{rod03,rod05}, as well.
 
The first application of the pn-cQRPA approach of Ref. \cite{rod03} to description of the 
$\beta\beta$-decay observables in several nuclei 
has been given recently in Ref.~\cite{rod07}. Realistic (zero-range) forces have been used in the p-p channel to describe 
the nucleon pairing within the BCS model realized on a rather large discrete+quasidiscrete s-p basis.

The pn-cQRPA approach of Refs. \cite{rod03,rod07} is further extended here by 
using a modern version of the phenomenological isoscalar mean field 
(including the spin-orbit term) deduced in Ref. \cite{igash06} from the isospin-selfconsistent 
analysis of experimental single-quasiparticle spectra in double-closed-shell 
nuclei. In the present contribution we give a brief overview of the approach and its 
applications to description of different GT strength functions for the pairs of 
nuclei $^{116}$Cd-Sn and $^{130}$Te-Xe.

\section{Coordinate representation of the pn-cQRPA equations and GT strength 
functions}

In formulation of a version of the pn-cQRPA we follow Ref. \cite{rod03}, where the 
pn-dQRPA equations originally written in terms of the forward $X^{(-)}_s$
and backward amplitudes $Y^{(-)}_s$ are transformed in equivalent equations 
for the 4-component radial transition density $\rho^{(-)}_i(s,r)$.
The latter is defined in terms of the $X$, $Y$ amplitudes by Eqs. (39), (40) of Ref. \cite{rod03} 
and related to the GT$^{(-)}$ excitations having the wave functions $|1^+\mu,s\rangle$ 
and energies $\omega_s$.
The expression for the transition density $\rho^{(+)}_i(s,r)$ related to 
GT$^{(+)}$ excitations follows from that for $\rho^{(-)}_i(s,r)$ by the 
substitution $p\leftrightarrow n$.
Hereafter, we often use the notations of Ref.~\cite{rod03} (except for the energies) and 
refer to some equations from this reference. 
Limiting ourselves in this contribution to description of the GT transitions only, we 
omit the quantum numbers indices $J=S=1$, $L=0$ for spin-monopole excitations. 
The spin-angular variables in all expressions are separated out as well.

The pn-dQRPA solutions $\omega_s$ are related to the excitation energies 
$E^{(\mp)}_{x,s}$ measured from the ground-state energy $E_0(Z\pm 1,N\mp 1)$ of 
the corresponding daughter nuclei as:
\be
\omega_s\pm(\mu_p-\mu_n)=\omega^{(\mp)}_s=E^{(\mp)}_{x,s}+Q^{(\mp)}_b.
\label{1}
\ee
Here, $\mu_{p(n)}$ is the chemical potential for the proton (neutron) 
subsystem found from the known BCS equations, 
$Q^{(\mp)}_b={\cal E}_b(Z,N)-{\cal E}_b(Z\pm 1,N\mp 1)$ are the total 
binding-energy differences, $\omega^{(\mp)}_s$ are the excitation energies 
measured from $E_0(Z,N)-\sum_a m_a=-{\cal E}_b(Z,N)$ 
($m_a$ is the nucleon mass).
The energies $\omega^{(\mp)}_s$ are usually described 
by a model Hamiltonian.

The system of equations for $\rho^{(-)}_i(s,r)$ (Eq.(41) of Ref. \cite{rod03}) 
contains the explicit expression for the $4\times 4$ matrix of the free 
two-quasiparticle propagator $A^{(-)}_{ik}(r,r',\omega_s)$ (Eq.(43) 
of Ref. \cite{rod03}; $A^{(+)}_{ik}=A^{(-)}_{ik}(p\leftrightarrow n)$).
These propagators are the main quantities in description of charge-exchange 
excitations within the pn-QRPA.
In particular, in terms of $A_{ik}$ one can formulate a Bethe-Salpeter-type 
equation for the effective propagator $\tilde{A}_{ik}(r,r',\omega_s)$ \cite{rod07}. 
The spectral expansion of $\tilde{A}_{ik}$ in terms of $\rho_i(s,r)$ allows one 
to express the pn-QRPA strength functions in terms of the effective propagator, 
or, equivalently, in terms of the 4-component effective fields \cite{rod03}. 
Some relevant formulas are shown below. 

The GT$^{(\mp)}$ strength functions, corresponding to the external fields (probing 
operators) $\hat{V}^{(\mp)}_{\mu}=\sum_aV^{(\mp)}_{\mu}(a)$, 
$V^{(\mp)}_{\mu}=\sigma_{\mu}\tau^{(\mp)}$, are defined as follows:
\be
S^{(\mp)}(\omega)=\sum_{s}|\langle 1^+,s\|\hat{V}^{(\mp)}\|0\rangle|^2\delta(\omega-\omega_s)
\label{2}
\ee
with GT strengths $B^{(\mp)}_s(GT)=|\langle 1^+,s\|\hat{V}^{(\mp)}\|0\rangle|^2$.
The strength function $S^{(-)}(\omega)$ can be expressed in terms of the corresponding 
effective field $\tilde{V}^{(-)}_{i[1]}(r,\omega)$, which is different from the external one
$V^{(-)}_i(r)=\delta_{i1}$ due to the residual interaction~\cite{rod03}:
\be
S^{(-)}(\omega)=-{{3}\over{\pi}}Im\sum_i\int A^{(-)}_{1i}(r,r',\omega)\tilde{V}^{(-)}_{i[1]}(r',\omega)drdr',
\label{3}
\ee
\be
\tilde{V}^{(-)}_{i[1]}(r,\omega)=\delta_{i1}+{{F^{(1)}_i}\over{4\pi r^2}}\sum_k
\int A^{(-)}_{ik}(r,r',\omega)\tilde{V}^{(-)}_{i[1]}(r',\omega)dr'.
\label{4}
\ee
The residual interaction here is supposed to be of zero-range type 
with intensities 
$F^{(1)}_i$: $F^{(1)}_1=F^{(1)}_2=2G'$, $F^{(1)}_3=F^{(1)}_4=G_1$.
For the $0^+$ p-h and p-p channels the corresponding strengths are: $F^{(0)}_1=F^{(0)}_2=2F'$ 
and $F^{(0)}_3=F^{(0)}_4=G_0$, respectively.
Dimensionless values $g'=G'/C$, $f'=F'/C$, ($C=300\ MeV\cdot fm^3$) are the well-known 
Landau-Migdal p-h strength parameters.
The same parameterization we use for the p-p interaction strengths: $g_1=G_1/C$, $g_0=G_0/C$. 
For calculation of $S^{(+)}(\omega)$ one can use Eqs. (3), (4) with substitution 
$p\leftrightarrow n$ \cite{rod03}.
An alternative way is based on the symmetry properties of $A_{ik}$: $A^{(+)}_{11}=A^{(-)}_{22}$. 
As a result, we get the expression for $S^{(+)}(\omega)$ in terms of $A^{(-)}_{ik}$ and 
$\tilde{V}^{(-)}_i$. 
This expression is obtained from Eqs. (3), (4) with the substitution $1\rightarrow 2$.

The nuclear GT$^{(-)}$ amplitude for $2\nu\beta\beta$-decay into the ground 
state $|0'\rangle$ of the product nucleus $(N-2,Z+2)$ is given by the expression:
\be
M^{2\nu}_{GT}=\sum_{s}{{\langle 0'\|\hat{V}^{(-)}\|1^+,s\rangle 
\langle 1^+,s\|\hat{V}^{(-)}\|0\rangle}\over{\bar{\omega}_s}},
\label{5}
\ee
where 
$\bar{\omega}_s=E_s-{{1}\over{2}}(E_0+E_{0'})=E_{x,s}+{{1}\over{2}}(Q^{(-)}_b+{Q_b^{(+)}}')$.
To calculate $M^{2\nu}_{GT}$ within the pn-QRPA, the vacua $|0\rangle$ and $|0'\rangle$ should be identified. 
As a result of such identification, one has $\bar{\omega}_s={{1}\over{2}}(\omega^{(-)}_s+{\omega_s^{(+)}}')\approx\omega_s$,
in accordance with Eq.~(1). 

The amplitude (5) can be expressed in terms of a ``non-diagonal" GT$^{(-)}$ strength 
function $S^{(--)}(\omega)$:
\be
M^{2\nu}_{GT}=\int \omega^{-1}S^{(--)}(\omega)d\omega,
\label{7}
\ee
where $S^{(--)}(\omega)$ is defined as follows:
\be
S^{(--)}({\omega})=\sum_{s}\langle 0'\|\hat{V}^{(-)}\|1^+,s\rangle
\langle 1^+,s\|\hat{V}^{(-)}\|0\rangle\delta({\omega}-\bar\omega_s).
\label{6}
\ee
The corresponding pn-QRPA expression for $S^{(--)}$ is:
\be
S^{(--)}(\omega)=-{{3}\over{\pi}}Im\sum_i\int A^{(-)}_{2i}(r,r',\omega)\tilde{V}^{(-)}_{i[1]}(r',\omega)drdr'.
\label{8}
\ee

An alternative expression for $M^{2\nu}_{GT}$ is obtained in terms of the ``non-diagonal" static polarizibility~\cite{rod07}:
\be
M^{2\nu}_{GT}=-{{3}\over{2}}\sum_i\int A^{(-)}_{2i}(r,r',\omega=0)\tilde{V}^{(-)}_{i[1]}(r',\omega=0)drdr'.
\label{9}
\ee
Decomposition of the amplitude (5) into two terms~\cite{rod05}
\be
M^{2\nu}_{GT}=(M^{2\nu}_{GT})'+\bar{\omega}^{-2}_{GTR}EWSR^{(--)},
\label{10}
\ee
\be
EWSR^{(--)}=\sum_{s}\bar\omega_s\langle 0'\|\hat{V}^{(-)}\|1^+,s\rangle 
\langle 1^+,s\|\hat{V}^{(-)}\|0\rangle,
\ee
where $\bar{\omega}_{GTR}$ is the energy of GT$^{(-)}$ giant resonance (GTR), allows us to 
clarify the sensitive dependence of $2\nu\beta\beta$-decay amplitude as a function of $g_{pp}$ 
(for details, see Ref. \cite{rod05}). 
The ``non-diagonal" energy-weighted sum rule $EWSR^{(--)}$ is straightforwardly expressed in terms of
the strength function $S^{(--)}$ of Eq. (6):
\be
EWSR^{(--)}=\int\omega S^{(--)}(\omega)d\omega,
\label{11}
\ee
again supposing the QRPA vacuum $|0'\rangle$ is identified with that of $|0\rangle$.

\section{Calculation of strength function within the pn-cQRPA}

Starting from the coordinate representation of the pn-dQRPA equations outlined above, we are able 
to take exactly into account the s-p continuum in the p-h channel and, therefore, to formulate a 
version of the pn-cQRPA.
The pairing problem is solved on a rather large basis of bound+quasibound 
proton and neutron s-p states within the present version of the model.
To take the s-p continuum into account, the following transformations of the expression for 
$A_{ik}(r,r',\omega)$ \cite{rod03} are done: 
(i) the Bogolyubov coefficients $v_{\lambda}$, $u_{\lambda}$ and the quasi-particle energies 
$E_{\lambda}$ are approximated by their non-pairing values $v_{\lambda}=0$, $u_{\lambda}=1$, and 
$E_{\lambda}=\varepsilon_{\lambda}-\mu$ for those s-p states ($\lambda$), which lie far above the chemical 
potential (i.e. $\varepsilon_{\lambda}-\mu\gg\Delta_{\lambda}$),
(ii) the Green function of the s-p radial Schr\"odinger equation 
$g_{(\lambda)}(r,r',\varepsilon)=\sum_{\varepsilon_{\lambda}}(\varepsilon-\varepsilon_{\lambda}+i0)^{-1}
\chi_{\lambda}(r)\chi_{\lambda}(r')$, which is calculated via the regular and irregular solutions of 
this equation, is used to perform explicitly the sum over the s-p states in the continuum. 
As a result, the properly transformed free two-quasiparticle propagator $A$ is obtained, upon which a corresponding 
version of the pn-cQRPA is based.

The solution of the pairing problem is simplified by using the ``diagonal" approximation for 
the p-p interaction for the $0^+$ neutral channel. 
In this approximation the nucleon-pair operators are assumed to be formed 
only from the pair of nucleons occupying the same s-p level $\lambda$.
The nucleon pairing is described with the use of the Bogolybov transformation with the gap parameter 
$\Delta_{\lambda}$ dependent on $\lambda$. 
The same number $N_{b+qb}$ of bound+quasibound states forming the basis of the BCS 
problem is used for both the neutron and proton subsystems. 
These numbers are shown in Table 1 for nuclei in question. 
In evaluation of total binding energies within the model (that is necessary to evaluate the pairing 
energies ${\cal E}_{pair}$) the blocking effect for odd nuclei is taken into account. 
In description of the nucleon pairing, different values of the p-p interaction strength parameters 
$g_{0,n}$ and $g_{0,p}$ for the neutron and proton subsystems are used.
These values are found from comparison of the calculated and experimental pairing energies for 
nuclei under consideration (Table 1). 

\begin{table}[ht]
\caption{The phenomenological mean field parameters ($U_0$, $U_{SO}$ and $a$), singlet and triplet p-h and p-p interaction 
strengths ($f'$, $g'$, $g_{0}$, $g_{pp}$) used in calculations.}
\begin{tabular}{|c|c|c|c|c|c|c|c|c|c|c|}     
\hline
Pair of nuclei         & $U_0$, MeV & $U_{SO}$, MeV$\cdot$fm$^2$ & $a$, fm  & $f'$ & $g_{0,n}$ & $g_{0,p}$ & $N_{b+qb}$ & $g'$ & $g_{pp}$ \\
\hline
$^{116}$Cd-$^{116}$Sn & 51.62      &  34.08                  & 0.618 & 1.06 & 0.388     &  0.333    &    22      & 0.77  &   1.0    \\
\hline
$^{130}$Te-$^{130}$Xe & 51.74      &  34.025                 & 0.628 & 1.09 & 0.356     &  0.364    &    22      & 0.88  &   0.99   \\
\hline
\end{tabular}
\end{table}

The mean field consists of the phenomenological isoscalar part (including the spin-orbit term) along 
with the isovector and Coulomb part (Eq. (1) of Ref. \cite{rod03}).
The parameterization of the Woods-Saxon-type isoscalar part contains two strength ($U_0$, $U_{SO}$) 
and two geometrical ($r_0$, $a$) parameters \cite{igash06}.
The mean field isovector part (the symmetry potential) is calculated in an isospin-selfconsistent way 
(Eqs. (7), (35) of Ref. \cite{rod03}) via the neutron-excess density and Landau-Migdal strength parameter $f'$. 
The mean Coulomb field is also calculated selfconsistently via the proton density. 
All densities are calculated with taking into account the nucleon pairing. 
Five above-listed model parameters found in Ref. \cite{igash06} for a number of double-closed-shell nuclei  are 
properly interpolated for nuclei under consideration (see Table 1; $r_0=1.27$ fm is taken for all nuclei).

The values of the Landau-Migdal strength $g'$ listed in Table 1 are obtained
by fitting the experimental GTR energy 
in calculations of the GT$^{(-)}$ strength function. The p-p interaction strength $g_1$ 
(or its relative value $g_{pp}=2g_1/(g_{0,n}+g_{0,p})$) is considered as a free parameter. 
It can be adjusted to reproduce the experimental $M^{2\nu}_{GT}$ value (the corresponding values are listed 
in the last column of Table 1).

Considering the pair $^{116}$Cd-$^{116}$Sn, the GT$^{(-)}$ strength 
distribution calculated within the pn-dQRPA for the transition $^{116}$Sn$\to ^{116}$Sb is shown 
in Fig. 1a (a small imaginary part is added to the s-p potential). 
To compare the calculation results with the $^{116}$Sn($^3$He,t) 
experimental data of Ref. \cite{pham95}, five centroids of the energy, $E_{x,i}$, and their strength 
$x_i$ relative to the one of the GTR are evaluated (Fig. 1b).
The value $g'=0.77$ allows to reproduce the experimental GTR energy in the calculation.
The GT$^{(-)}$ strength distribution is almost insensitive to the $g_{pp}$ value ($g_{pp}=1.0$ is taken in 
the calculation). 
The GT$^{(+)}$ strength distribution for the transition $^{116}$Sn$\to ^{116}$In is found more sensitive to $g_{pp}$.
Only one $1^+$ state with $B^{(+)}(GT)=0.47$
corresponding to the $1g^p_{9/2}\rightarrow 1g^n_{7/2}$ transition into the $^{116}$In ground 
state, is found in the calculation within the interval $E_x<5\ MeV$.
This weak transition is allowed due to the neutron pairing in $^{116}$Sn.
In the $^{116}$Sn(d,$^2$He) experiments four $1^+$ states in $^{116}$In were found within the interval 
$E_x\le 3\ MeV$ with total strength $\sum_iB_i^{(+)}(GT)=0.66$ \cite{rak05}.
Population of the $1^+$ states in $^{116}$In has also been studied in the $^{116}$Cd(p,n)-reaction \cite{sas07}.
The result $B^{(-)}(GT)=0.26\pm 0.02$ for excitation of the $^{116}$In ground state is only 
available now.
Within the interval $E_x\le 3\ MeV$ the calculated GT$^{(-)}$ strength distribution in 
$^{116}$In exhibits one $1^+$ state, corresponding to the back-spin-flip transition 
$1g^n_{7/2}\rightarrow 1g^p_{9/2}$ into the $^{116}$In ground state with the value 1.05 $B^{(-)}(GT)$. 
(for $^{116}$Sn-$^{116}$Sb this transition is Pauli blocked). 
The $2\nu\beta\beta$-decay amplitude for the decay $^{116}$Cd$\to^{116}$Sn can barely be evaluated within the pn-QRPA 
because the proton shell is closed in $^{116}$Sn.

Coming to the pair $^{130}$Te-$^{130}$Xe, the value $g'=0.88$ is found 
in the calculation by fitting the experimental GTR energy in $^{130}$I~\cite{mad89}.
Then the amplitude $M^{2\nu}_{GT}$ (6) (or (9)) and its decomposition (10) are calculated, as 
a function of $g_{pp}$ (Fig. 2). The corresponding experimental value 
$(M^{2\nu}_{GT})^{exp}=0.03$ MeV$^{-1}$ \cite{bar06} can be reproduced in the calculation at $g_{pp}=0.99$.
The $2\nu\beta\beta$-decay strength function $\omega^{-1}S^{(--)}(\omega)$ is calculated 
for this value of $g_{pp}$ (Fig. 3a) along with the corresponding running sum 
$M^{2\nu}_{GT}(\omega)=\int^{\omega}{\omega'}^{-1}S^{(--)}(\omega')d\omega'$ (Fig. 3b).
Figs. 2 and 3 illustrate how the $M^{2\nu}_{GT}$ value for the decay $^{130}$Te$\to^{130}$Xe is formed.
In particular, as one sees in Fig.3, the experimental studies of $B^{(+)}(GT)$ are not always sufficient
for understanding partial contributions to $M^{2\nu}_{GT}$. 
The reason is that the intermediate states having a relatively 
large excitation energy and very small $B^{(+)}(GT)$ value (like the GTR) can nonetheless 
play essential role in formation of the $2\nu\beta\beta$-decay amplitude.

In conclusion, an isospin-selfconsistent version of the pn-cQRPA has been outlined and some its applications to 
description of charge-exchange excitations in open-shell spherical nuclei are presented.
Although only general features of the low-energy strength distributions can be described within the approach, 
it seems applicable to analysis of $\beta\beta$-decay observables.

This work is supported  in part by RFBR (grant 06-02-016883-a) (S.I. and M.U.), NWO (M.U.), DFG (grant FA67/28-2) 
and EU ILIAS project (contract RII3-CT-2004-506222).



\begin{figure}
\ \hskip-3cm\includegraphics[scale=0.37]{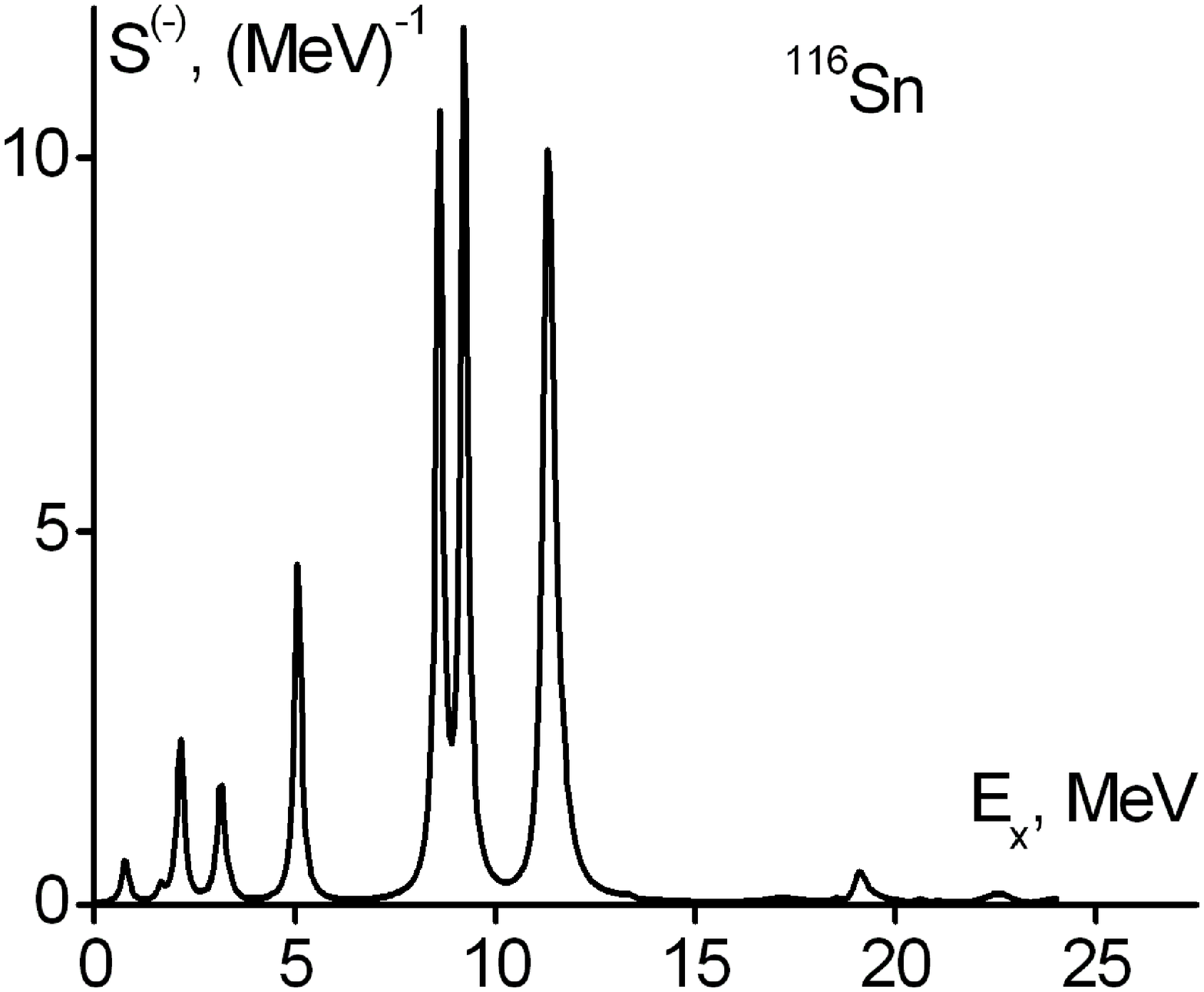}\ \hskip-1cm\includegraphics[scale=0.37]{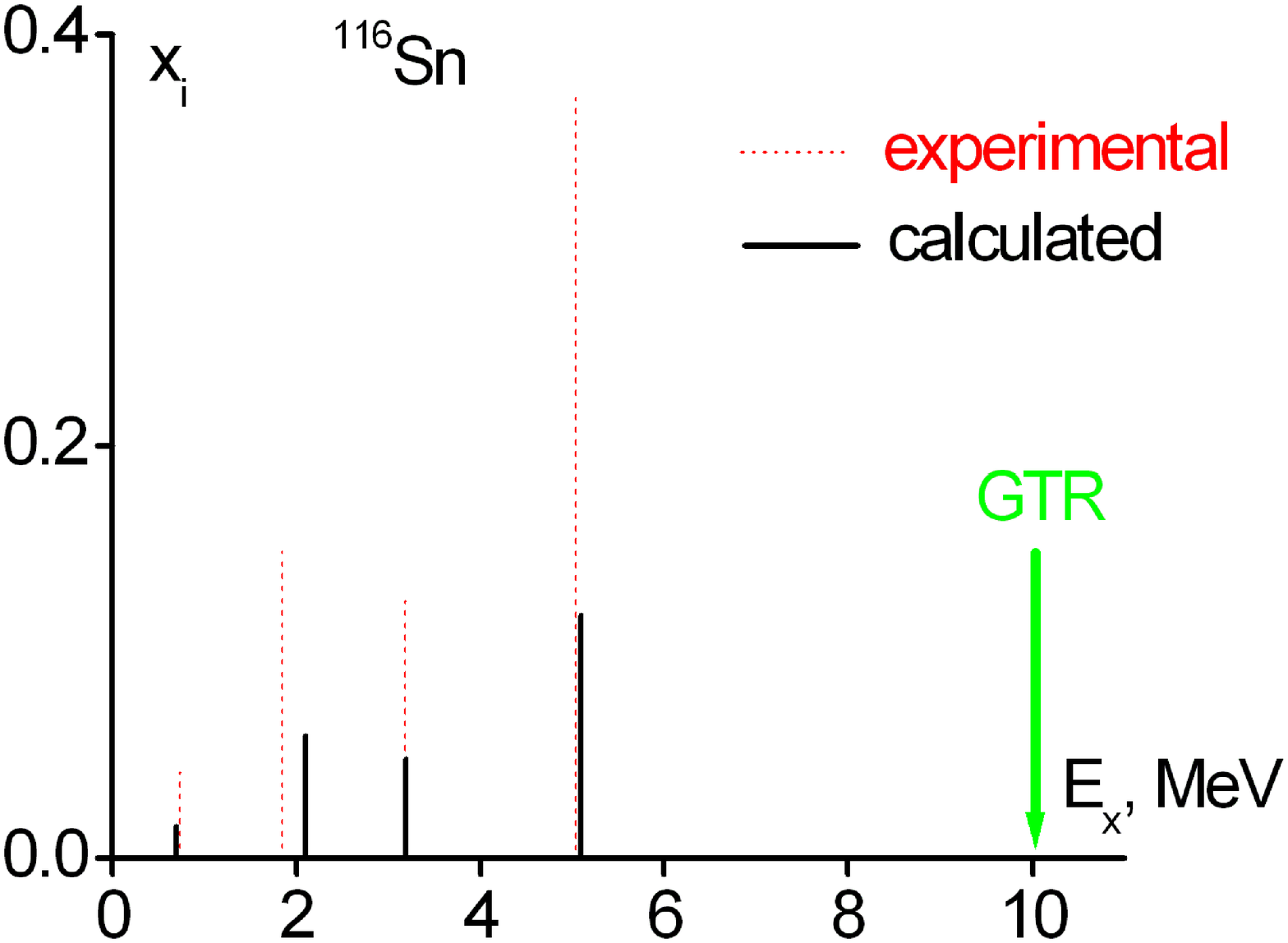}
\caption{The GT$^{(-)}$ strength function for $^{116}$Sn-$^{116}$Sb 
(a) and the relative (with respect to the GTR) strength of the 
low-energy $1^+$ peaks calculated within pn-cQRPA (b). The corresponding experimental 
data are taken from Ref. \cite{pham95}}
\end{figure}

\

\begin{figure}
\begin{center}
\includegraphics[scale=0.5]{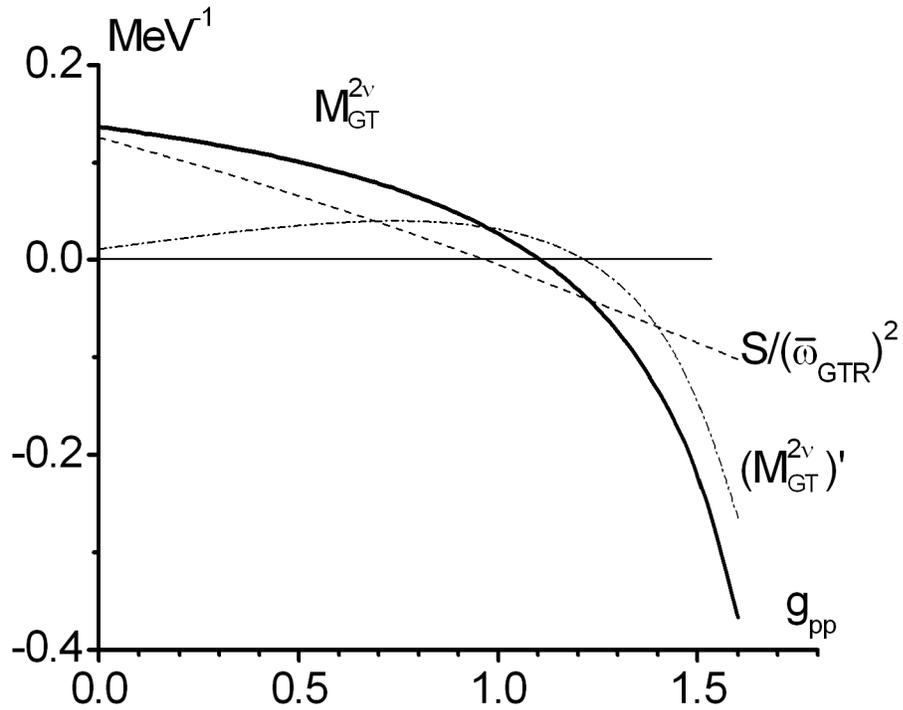}
\end{center}
\caption{The calculated amplitude of $^{130}$Te $2\nu\beta\beta$-decay as a function of $g_{pp}$. 
Decomposition of Eq. (10) is also shown.}
\end{figure}

\begin{figure}
\ \hskip-2cm \includegraphics[scale=0.37]{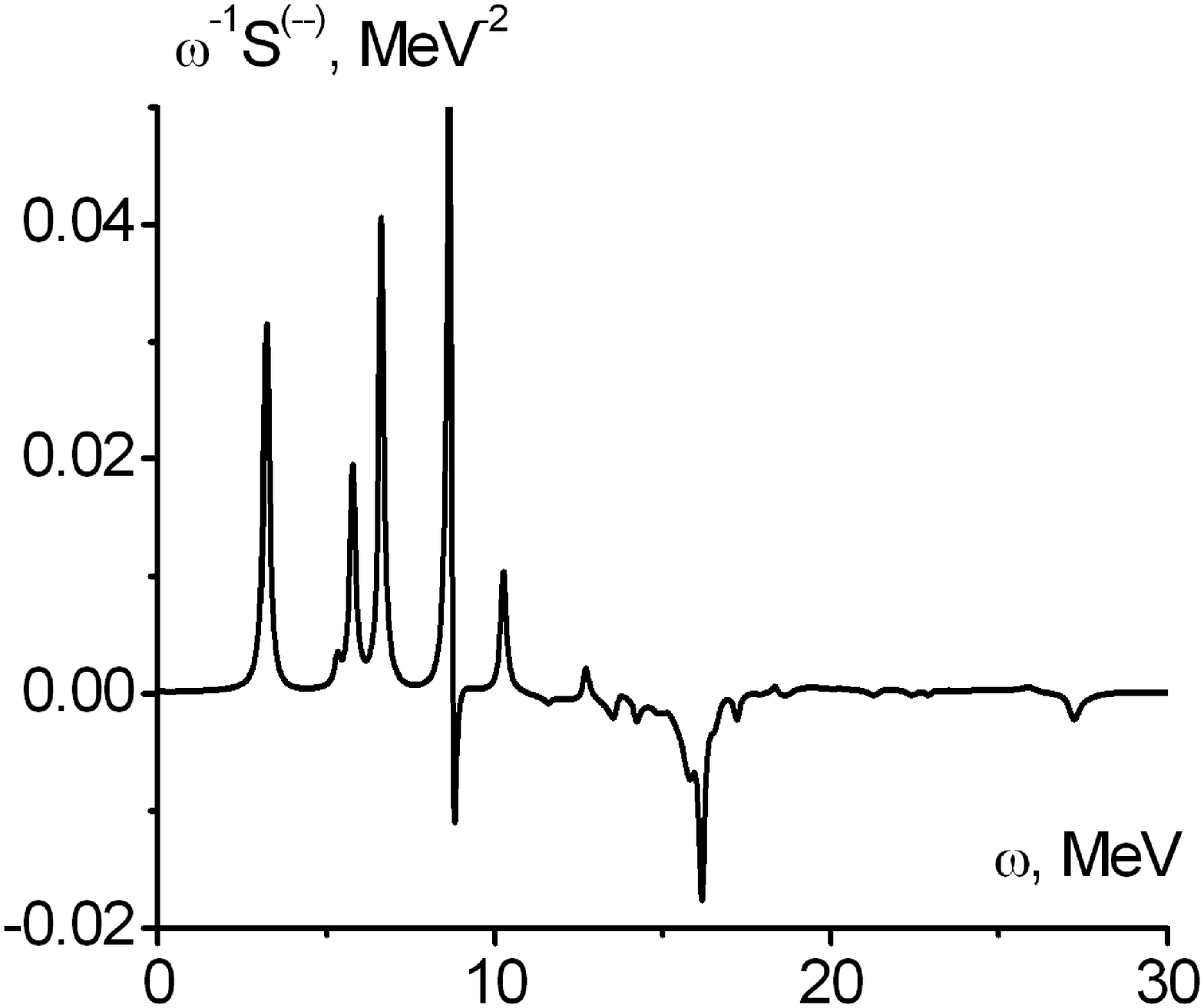}\ \hskip-1cm\includegraphics[scale=0.37]{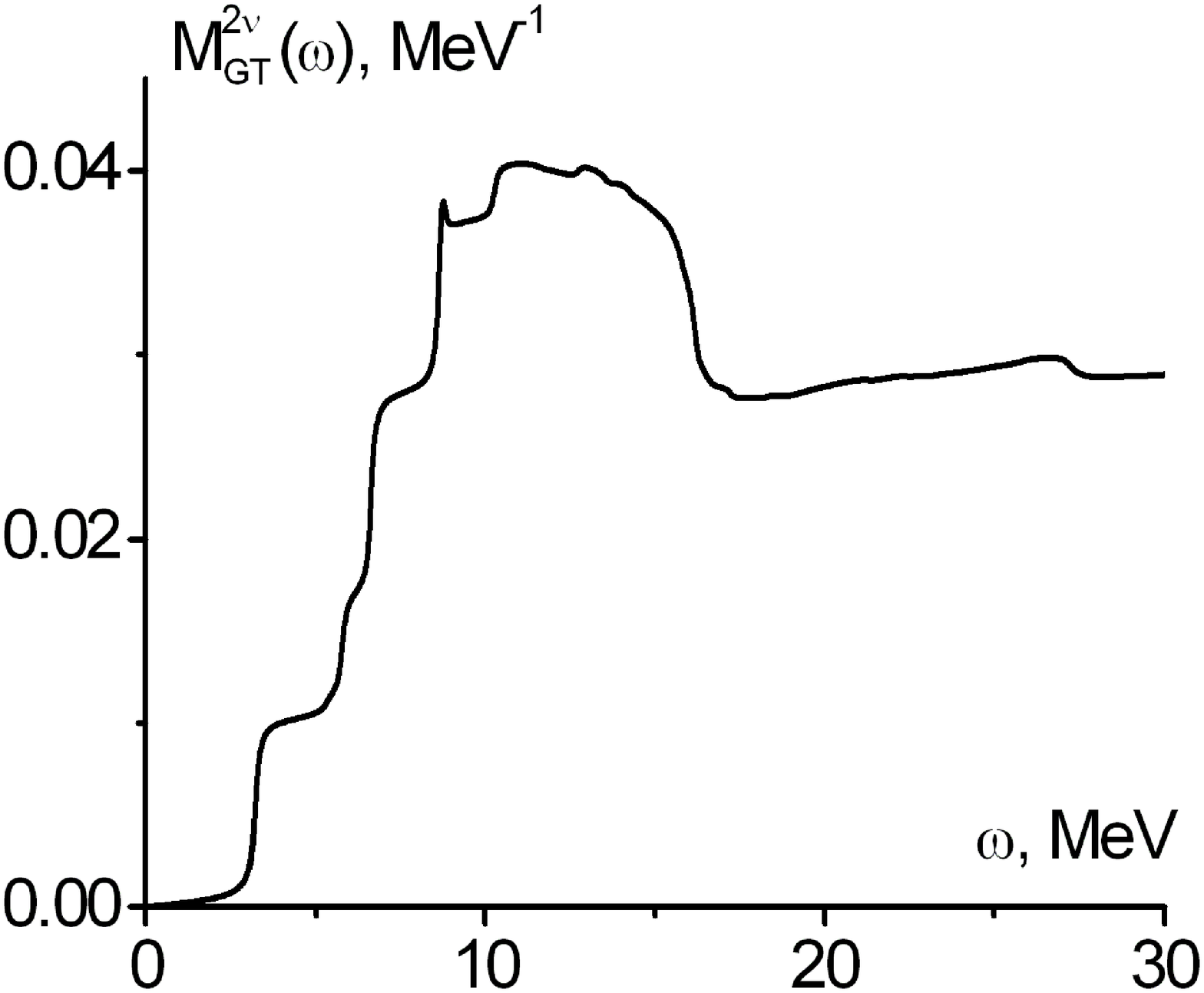}
\caption{The GT $2\nu\beta\beta$-decay strength function (a) and the running sum (b) calculated for 
$^{130}$Te at $g_{pp}=0.99$}
\end{figure}

\end{document}